\documentclass[12pt]{article}
\usepackage{amsfonts}
\usepackage{amssymb}
\usepackage{amscd}
\def\beq{\begin{equation}}
\def\eeq{\end{equation}}
\def\bea{\begin{eqnarray}}
\def\eea{\end{eqnarray}}
\def\ba{\begin{array}}
\def\ea{\end{array}}
\def\bi{\begin{itemize}}
\def\ei{\end{itemize}}
\def\i{\item}

\def\lrb{\left(}
\def\rrb{\right)}

\def\vsj{\vspace{1mm}}
\def\vsd{\vspace{2mm}}
\def\vst{\vspace{3mm}}
\def\vsc{\vspace{4mm}}
\def\vsp{\vspace{5mm}}

\def\hsp{\hspace{5mm}}

\def\R{\Bbb R}

\def\C{\Bbb C}

\def\N{\Bbb N}

\def\al{\alpha}

\def\noi{\noindent}
\def\bt{\begin{theorem}}
\def\et{\end{theorem}}
\def\noi{\noindent}
\def\qk{\hat{q}\,}
\def\pk{\hat{p}\,}
\def\hk{\hat{h}\,}
\def\rk{\hat{\rho}\,}
\def\ik{\hat I\,}
\def\ooo{{\cal{O}}}
\def\sss{\cal{{S}}}
\def\bbb{\cal{{B}}}
\def\bs{{\bf S}}
\def\htri{{\cal U}\,({\cal H}_3 )}
\def\alg{{\cal U}}
\def\nn{\nonumber}

\def\R{{\mathbb R}}
\def\N{{\mathbb N}}
\def\qed{{\ }\hfill$\diamondsuit$}
\def\htnm{\htri_{n,m}}

\def\anm{\hat A_{n,m}}

\def\om{{\Omega}_{n,m}^2}

\def\ed{\end{document}}

\begin{document}
\baselineskip18pt
\title{Weyl ordering rule and new Lie bracket of quantum mechanics}
\author{Zoran Raki\'c,\thanks{e-mail:\,
zrakic@matf.bg.ac.yu}\ \ \ Faculty of Mathematics, \\ P.O. Box
550, 11000
Belgrade, Serbia, \\
and \\
Slobodan Prvanovi\'c,\thanks{e-mail:\, prvanovic@phy.bg.ac.yu }
 \ \ \ Institute of Physics, \\ P.O. Box 57, 11080 Belgrade, Serbia. }

\date{}
\maketitle
\begin{abstract}
\noi The product of quantum mechanics is defined as the ordinary
multiplication followed by the application of superoperator that
orders involved operators. The operator version of Poisson bracket
is defined being the Lie bracket which substitutes commutator in
the von Neumann equation. These result in obstruction free
quantization, with the ordering rule which coincides with Weyl
ordering rule.
\end{abstract}

$$ \ $$
$$ \ $$
$$ \ $$
$$ \ $$

\pagebreak
\noi {\Large {\bf{ I\ \ \,  Introduction}}} \vspace{4mm}

\noi Quantization may roughly be thought as transition from
classical to quantum description of a system. What is its precise
definition is not generally agreed since many of related questions
are still opened. With different motivations, quantization was
approached in different manners, see \cite{1} and references
therein. Even in this case, the  ordering problem is not trivial.
The most important problem regarding the quantization of
${\R}^{2n}$ is in that it is not known how should the algebraic
and Lie algebraic multiplications of quantum mechanics be realized
in unambiguous and consistent way. \vspace{1mm}

\noi In ~\cite{2,3} one finds a short review of different
propositions of symmetrized product. In ~\cite{3} it was shown
that there given propositions for symmetrized product start to
differ for quadratic monomials in $\hat q$ and $\hat p$. In
~\cite{4} there is a critical discussion of many ordering rules
introduced in there cited references. In ~\cite{2,5,6,7} and
references therein, it was found that algebraic and Lie algebraic
structures of quantum mechanics are interrelated in such a way
that obstructions result in quantization which manifest themselves
through some contradictions in a formalism. A detailed derivation
of the Lie bracket of quantum mechanics in relation to
quantization one can find in ~\cite{8}. (Needless to say, as the
standard choice of the  Lie bracket of quantum mechanics appears
the commutator divided by $i\hbar$.) Because of the mentioned
contradictions, one can conclude, see ~\cite{7}, that the problem
of quantization is impossible or, as was noticed in ~\cite{1},
that some subtler symmetrization rule is necessary. \vspace{1mm}

\noi  In this article we shall define the symmetrized product of
quantum mechanics. Then, by using this product, we shall be able
to propose new Lie bracket of quantum mechanics. It will be the
operator version of Poisson bracket. Since in the classical
mechanics this bracket is used in equation of motion - the
Liouville equation, we shall propose the reformulation of
dynamical equation of quantum mechanics. It will be shown that the
von Neumann - Schr\"odinger equation can be seen as the operator
version of Liouville equation. In this way quantum and classical
mechanics will appear to be the same regarding the algebraic and
Lie algebraic aspect and the dynamical equation. The same
realization of these crucial elements of the theories will allow
us to propose the quantization of ${\R}^{2n}$ that is free of
obstruction. Finally,  we will show that the ordering rule given
by our quantization coincide with Weyl ordering rule. This fact is
proved in Theorem 5. \vspace{10mm}

\noi {\Large {\bf{ II\ \,  The symmetrizer and symmetrized
product}}}  \vspace{4mm}

\noi Let us introduce basic notations and notions. Let
$\om(\qk,\pk)=\om,$ be the permutation group (with repetition) of
$n$ examples of  $\qk$ and $m$ of $\pk.$ $\htri$ will denote the
universal enveloping algebra of Heisenberg algebra generated by
$\{ \qk,\pk, \ik \}$ and with the only nontrivial commutator
\beq\label{e21} [\,\qk,\pk\,]=i\,\hbar\, \hat I .\eeq If we
introduce, the operator $\hk=-\,i\,\hbar\, \hat I,$ then because
of (\ref{e21}) it is clear that the set\linebreak ${\bbb} = \{\,
\qk^n\,\pk^m\,\hk^p\,\mid\, n,m,p\in \N_{\,0}\, \} $ is a
PBW-basis of $\htri .$ We will use the following notations
$\hat\htri,$ for subspace of $\htri$ generated by $\pk$'s and
$\qk$'s;
 by $\hat\htri_{n,m}$ we will denote the subspace of $\htri$
generated by all possible monomials with $n$\, $\qk$'s and $m$
$\pk$'s, {\it i.e.}, an arbitrary expression of the form
$$\hat q ^{n_1}\,\, \pk ^{m_1}\,\, \cdots \,\, \hat q ^{n_s}\,\, \pk ^{m_s},
$$ where $n_i\geq 0,\ m_j\geq 0\ $ and $\ \sum_{i=1}^s n_i=n,\ \ \
\sum_{i=1}^s m_i=m. $ \vst

\noi {\bf Definition 1.}\ \ \  Let $\alg$ be a $\C-$associative
algebra with unit generated by $\{\pk,\qk,\hk,\hat \rho, {\partial
\hat \rho \over \partial \hat r }\}.$  {\bf Symmetrizer} $\bf S$
is a linear map, $\bs:\alg \longrightarrow \alg,$ defined as
follows: $\forall\, \hat A_{n,m}\in\hat\htri_{n,m}$  \bea
& & {\bf S}(\hat A_{n,m} ) =  \lrb \ba{c} n+m \\
        n\ea\rrb ^{-1}\,
 \sum_{\sigma(\qk,\pk)\, \in\, \Omega_{n,m}^2}
\sigma(\qk,\pk) \label{e22}\vspace{2mm}\\
& & {\bf S}(\anm \cdot \hat \rho )= \anm\cdot \rk, \qquad
 {\bf S}(\hat \rho \cdot \anm )= \hat \rho \cdot \anm ,\label{e23}\vst\\
& & {\bf S}(\anm \cdot {\partial \hat \rho \over
\partial \hat r } )=  \lrb \ba{c} n+m+1 \\
        n\ea\rrb ^{-1}
 \sum_{\sigma(\qk,\pk) \in \Omega_{n,m,1}^3}
\sigma(\qk,\pk,{\partial \hat \rho \over
\partial \hat r }) \label{e24},\vsd \\
& & {\bf S}({\partial \hat \rho \over
\partial \hat r } \cdot \anm ) =  \lrb \ba{c} n+m+1 \\
        n\ea\rrb ^{-1}
 \sum_{\sigma(\qk,\pk) \in \Omega_{n,m,1}^3}
\sigma(\qk,\pk,{\partial \hat \rho \over
\partial \hat r }) \nn .\vsd\eea

\noi {\it Remark.} (a) Because of (\ref{e21}) and (\ref{e22}), it
is clear that  ${\bf S}(\hat B)=0$, for every monomial $\hat B\in
\htri$ which has $\hk$ as a factor.\vspace{1mm}

\noi (b) From the above definition it is clear that $ \bs$
maps all $\hat\htnm $ onto $$  \lrb \ba{c} n+m \\
        n\ea\rrb ^{-1}\,
 \sum_{\sigma(\qk,\pk)\, \in\, \Omega_{n,m}^2}
\sigma(\qk,\pk).$$ \vsj

\noi (c) Since all different combinations of involved $\qk$'s and
$\pk$'s appear on the RHS of (\ref{e22}), the symmetrized product
of two symmetrized monomials, which are the Hermitian, obviously
will be invariant under the Hermitian conjugation.\vsj

\noi (d)  From the physical reasons  the action of operator $\bs$
on another elements of $\alg$ are not important for us (here).
\vsd

\noi Definition of symmetrizer enables us to introduce symmetrized
product in the following way. \vsj

\noi {\bf Definition  2.}\ \ \ For any two $\hat A, \hat B \in
\hat\htri, $ the symmetrized product is composition of ordinary
multiplication and application of symmetrizer:
\begin{equation}
\hat A \circ \hat B = {\bf S}(\hat A \cdot \hat B ).\label{e25}
\end{equation}

\noi {\sl Example.} The symmetrized product of the square of
operator of coordinate and the square of operator of momentum is:
\begin{equation}\label{e26}
\qk ^2 \circ \pk ^2 = {\bf S}(\qk ^2 \, \pk ^2 )= {1\over 6}\,(\qk
^2 \pk ^2 + \qk \pk\, \qk \pk + \qk \pk ^2 \qk + \pk\, \qk ^2 \pk
+ \pk\, \qk \pk\, \qk + \pk ^2 \qk ^2 ) .
\end{equation}
The formulas, specially (\ref{e26}) and more general (\ref{e22}),
are the most 'natural' (with respect to symmetry) generalization
of famous Dirac's symmetrized product given by ${1\over 2}\,\,
(\hat q\, \pk + \pk\, \hat q ).$  \vst

\noi {\bf Proposition 1.}\ \ Symmetrized product $\circ$ is a
commutative map.\vsj

\noi {\sl Proof:} It follows from
\begin{equation}\label{e27}
(\qk ^a \circ \pk ^b ) \circ (\qk ^c \circ \pk ^d )={\bf S}({\bf
S}(\qk ^a \cdot \pk ^b )\cdot {\bf S}(\qk ^c \cdot
 \pk ^d ))= \qk ^{a+c} \circ \pk ^{b+d} \,. \hspace{15mm} \mbox{\qed}
\end{equation}\vsj

\noi The ordered product of $f(\hat {\vec r})$ and $\hat {\vec p}$
has to be the one half of the anti-commutator of these two (this
appears in the Hamiltonian of charged particle in the
electromagnetic field), see ~\cite{10}. Since with our proposal of
symmetrized product we do not want to contradict the well-known
facts of standard quantum mechanics, we have to prove the
following proposition.\vst

\noi {\bf Proposition 2.}\ \ In the algebra $\alg$ the following
properties hold: \bea &\mbox{(a)}& \qquad {1\over 2}\,(\sum _{j=0}
^n c_j\, \qk ^j\, \pk + \pk \sum _{j=0} ^n c_j\, \qk
^j )=\sum _{j=0} ^n c_j\, \qk ^j \circ \pk\,, \label{e28}\vsd\\
&\mbox{(b)}& \qquad {\partial \over \partial \qk} \sum _{j,k\geq
0} c_{jk}\, \qk ^j \circ \pk ^k = \sum _{j,k\geq 0} j \, c_{jk} \,
\qk ^{j-1}
\circ \pk ^k ,\label{e29}\vsd\\
& \mbox{(c)} & \qquad {\partial \over \partial \pk} \sum _{j,k\geq
0} c_{jk}\, \qk ^j \circ \pk ^k = \sum _{j,k\geq 0} k \, c_{jk} \,
\qk ^j \circ \pk ^{k-1}. \label{e210}\eea \vsj

\noi {\sl Proof.}
 (a) The LHS can be transformed
into: $$ \sum _{j=0} ^n c_j\, \qk ^j\, \pk  - {i\hbar \over 2}\,
{\partial \over
\partial \qk}\sum _{j=0} ^n c_j\, \qk ^j .
$$
For the RHS it holds: \begin{small}\bea\nn & & \hspace{-15mm}\sum
_{j=0} ^n c_j {1\over {j+1}} \,(\pk\, \qk ^j + \qk \pk\, \qk
^{j-1} +\cdots + \qk ^{j-1} \pk\,
\qk + \qk ^j \pk )\vsd \\
& & \hspace{-13mm} = \  \sum _{j=0} ^n c_j\, {1\over j+1}\, (\qk
^j \pk -i\,\hbar\, j\, \qk ^{j-1} + \qk\, (\qk ^{j-1} \pk
-i\,\hbar\, (j-1)\, \qk ^{j-2}) +\cdots
+\qk ^{j-1}\nn\vsd \\
& &\hspace{-13mm} \times\, (\qk\, \pk -i\,\hbar )+ \qk ^j\, \pk
)=\sum _{j=0} ^n c_j\, \qk ^j\, \pk -i\, \hbar \sum _{j=0}
^n c_j\, {1\over {j+1}}\, (j + (j-1) + \cdots +1)\, \qk ^{j-1}\nn\vsd \\
&\hsp\hsp =& \sum _{j=0} ^n c_j\, \qk ^j \pk - {i\,\hbar \over
2}\, \sum _{j=0} ^n c_j\, j\, \qk ^{j-1}\nn = \sum _{j=0} ^n c_j\,
\qk ^j\, \pk - {i\, \hbar \over 2}\, {\partial \over\partial \qk}
\sum _{j=0} ^n c_j\, \qk ^j . \vsd\eea\end{small} Therefore, both
sides are equal.\vsj

\noi (b) Due to the linearity of partial derivations, it is enough
to prove this equation for monomials. In $\qk ^j \circ \pk ^k$
there are ${(j+k)\,!\over j\,!\,\, k\,!}$ different sequences in
the sum. Each sequence contains $j$ operators of coordinate and
$k$ operators of momentum. Partial derivation with respect to
$\qk$ produce $j$ new terms from each of these sequences, so there
are $j\,{(j+k)\,!\over j\,!\,\, k\,!}$ terms in the sum after this
derivation. Each of these new terms is the sequence of $j-1$
operators of coordinate and $k$ operators of momentum, as is
needed for $\qk ^{j-1} \circ \pk ^k$. Since the number of
different combinations of $j-1$ operators of coordinate and $k$
operators of momentum is less than $j\,{(j+k)\,!\over j\,!\,\,
k\,!}$, many of these new sequences are the same. Each of
${(j-1+k)\,!\over (j-1)\,!\,\,k\,!}$ different sequences needed
for $\qk ^{j-1} \circ \pk ^k$ appears $j+k$ times among the new
terms and in this way the multiplicative factor
\,${j\,!\,\,k\,!\over (j+k)\,!}$\,,\, standing in front of the sum
and coming from $\qk ^j \circ \pk ^k$ is regularized. So, the
proper multiplicative factor needed for $\qk ^{j-1} \circ \pk ^k$
is gained.

\vsj

\noi (c) Similarly to the previous case. \qed \vspace{10mm}

\noi {\Large {\bf{ III\ \,  The symmetrized Poisson
bracket}}}\vspace{4mm}

\noi Using the symmetrized product in the previous section we can
introduce the corresponding Poisson bracket. \vsd

\noi  {\bf Definition 3.} \ \ The \ {\bf sym\-me\-tri\-zed
Poi\-sson brac\-ket} of $\hat A,\, \hat B\in \hat\htri $ is given
by:
\begin{equation}\label{e311}
\{ \hat A , \hat B \} _{\bf S} =  {\partial \hat A \over
\partial \qk } \circ {\partial \hat B \over \partial \pk }
- {\partial \hat A \over \partial \pk } \circ {\partial \hat B
\over \partial \qk } .
\end{equation}\vsd

\noi The most important properties of symmetrized Poisson bracket
are content of the following\vst

\noi {\bf Proposition 3.} (a) The  symmetrized Poisson bracket is
the Lie bracket of quantum mechanical symmetrized observables.\vsj

\noi (b)  The symmetrized Poisson bracket is a derivative:
\begin{equation}\label{e312}
\{  \hat A , \hat B \circ \hat C \} _{\bf S} = \{ \hat A , \hat B
\} _{\bf S} \circ \hat C +  \hat B \circ \{  \hat A , \hat C \}
_{\bf S} .
\end{equation}\vsj

\noi {\sl Proof.}  That $\{ \ , \ \} _{\bf S}$ is linear holds due
to the fact that the partial derivations and application of
symmetrizer are linear operations. That it is anti-symmetric
follows from the commutativity of symmetrized product. The
confirmation of the Jacobi identity can rest on the analogy
between algebraic products of quantum and classical mechanics and
their relations with partial derivations. Each step of the
calculation in the case of operators $\qk ^{n'} \circ \pk ^{m'}$,
$\qk ^{n''} \circ \pk ^{m''}$ and $\qk ^{n'''} \circ \pk ^{m'''}$
has the corresponding one in the c-number case for $q^{n'} \cdot
p^{m'}$, $q^{n''} \cdot p^{m''}$ and $q^{n'''} \cdot p^{m'''}$
which satisfy the Jacobi identity. Due to the linearity of $\{ \ ,
\ \} _{\bf S}$, this identity holds for polynomials and analytical
functions of $\qk$ and $\pk$ with these two operators multiplied
according to the symmetrized product. \vsj

\noi (b) Confirmation is trivial due to the one-to-one relation
with the c-number case. \qed\vst

\noi Since the Poisson bracket is crucial part of the Liouville
equation, it does not come as surprise that we want to consider
the question whether it is possible to reexpress the dynamical
equation of quantum mechanics. But, before addressing this topic,
let us remark that the general state of quantum mechanical system
$\hat \rho $ can be expressed via $\hat q$ and $\hat p$. Details
are given in Appendix A. If it is seen as $\rho (\hat q ,\hat p ,
t)$, this operator can be derived with respect to $\hat q$ and
$\hat p$ directly, as it is done in the case of probability
distribution $\rho $ that describes state of classical mechanical
system.\vst

\noi {\bf Theorem 4.} Let $\hat H=\sum _i c_i\, \qk ^{n_i} \circ
\pk ^{m_i}$ be a Hamiltonian, then the following relations hold
 \bea & & \mbox{(a)}\qquad\{ \hat H , \hat
\rho \} _{\,\bf S} =
{1\over i\hbar }\, [\, \hat H , \hat \rho\, ], \hspace{75mm}\vsd \label{e313}\\
& & \mbox{(b)}\qquad {\partial \hat \rho \over \partial t }= \{
\hat H , \hat \rho \} _{\, \bf S},\label{e314}\hspace{75mm}
\eea\vsj

\noi {\sl Proof.} (a) Due to the linearity of symmetrized Poisson
bracket and commutator, this equation holds if it holds for $\hat
H =\qk ^n \circ \pk ^m$. The LHS of (\ref{e313}) in the case of
monomial, after partial derivations of $\hat H$ and
multiplications, becomes:
\begin{small}\begin{equation}\label{e315} {n!\, m! \over (n+m)!}
(\qk ^{n-1} \pk ^m {\partial \hat \rho \over
\partial \pk} + \cdots + {\partial \hat \rho \over
\partial \pk} \pk ^m \qk ^{n-1} - (\qk ^n \pk
^{m-1} {\partial \hat \rho \over \partial \qk} + \cdots +
{\partial \hat \rho \over \partial \qk} \pk ^{m-1} \qk ^n )).
\end{equation}\end{small}
After substituting ${\partial \hat \rho \over \partial \pk}$ with
${1\over i\,\hbar}\, [\, \qk , \hat \rho\, ]$ and ${-\partial \hat
\rho \over
\partial \qk}$ with ${1\over i\,\hbar}\, [\, \pk , \hat \rho\, ]$, (\ref{e315})
 can be simplified. Some terms in this expression are of
 the form $\hat
A \qk\, (\qk\, \hat \rho )\, \qk \hat B$, where $\hat A,\,\hat B$
represent (different) sequences of $\qk$'s and $\pk$'s and $(\qk\,
\hat \rho )$ means that these two come from the commutator $[\,
\qk , \hat \rho\, ]$. In transformed (\ref{e315}) terms $-\hat A\,
\qk\, \qk\, (\hat \rho\,  \qk )\, \hat B$, where the minus sign
comes from the commutator, certainly appear as well. So, these
terms mutually cancel each other. This holds for all other forms
of terms except for those where $\hat \rho$ stands at the
beginning or at the end of the sequence. Consequently,
(\ref{e315}) is equal to:
$$
{1\over i\hbar }\, {n!\, m! \over (n+m)!}\, ((\qk ^n \pk ^m +
\cdots + \pk ^m \qk ^n )\, \hat \rho  - \hat \rho\, (\qk ^n \pk ^m
+ \cdots + \pk ^m \qk ^n )),
$$
which is nothing else than the RHS of (\ref{e313}) for the
considered monomial.\vsj

\noi (b) Directly follows from (a).\qed \vsd

\noi {\it Remark.} The equation (\ref{e314}) is {\bf the dynamical
equation of quantum mechanics}. Obviously, von Neumann equation
(\ref{e314}) is symmetrized version of the Liouville equation.\vsd

\noi From the above given, it follows that one can propose
quantization which is, we believe, unambiguous, {\it i.e.},
obstruction free {\it in toto}.  \vsd

\noi {\bf Definition 4.}\ \ Let  Let the algebra of variables of a
classical mechanical system be generated by 1, $q$ and $p$, then
quantization is transition to operator formulation defined in the
following way: \bea 1, \ q , \ p \
&\stackrel{}{\longrightarrow}&\, \hat h , \
\qk , \ \pk ,\label{e316}\vsd\\
\cdot  &\stackrel{}{\longrightarrow}&
\circ\,,\label{e317}\vsd\\
\{\ ,\ \} &\stackrel{}{\longrightarrow}&  \{ \ ,\  \} _{\bf
S}\,.\label{e318} \eea\vst\

\noi {\sl Remark.} The Hamiltonian function of classical
mechanical system:  \bea\nn H(q,p)=\sum _i c_i\, q^{a_i} \cdot
p^{b_i}\,,\eea is mapped in the Hamiltonian \bea\nn H(\qk , \pk )=
\sum _i c_i\, \qk ^{a_i} \circ \pk ^{b_i}\,,\eea of quantum
mechanical system and dynamical equations of these theories are
related in the following way: \begin{small}\bea {\partial \rho
(q,p,t) \over
\partial t} = \{ H(q,p) , \rho (q,p,t) \}
&\stackrel{}{\longrightarrow}& {\partial \rho (\qk , \pk ,t) \over
\partial t}= \{ H(\qk , \pk ) , \rho (\qk , \pk , t )
\} _{\bf S}\,.\nn \label{e319} \eea\end{small}\vspace{5mm}

\noi {\Large {\bf{IV\ \ \, Symmetrized product and}}} \vspace{3mm}

\hspace{9mm}{\Large {\bf{Weyl ordering rule}}}\vspace{4mm}

\noi It is well known that the Weyl ordering rule is given by the
following formula $${\ooo} (q^n\,p^m)  =  \frac{1}{2^{\,n}}
\sum_{i=0}^n \, \qk^{\, n-i}\, \pk^{\, m}\,\qk^{\, i},$$ and also
it is known that Weyl quantization is obstruction free. \vsj

\noi In this section we will prove that our symmetrized product
and Weyl ordering rule are the same. This fact explains why Weyl
quantization is so symmetric. More precisely we will prove,

\vspace{3mm} \pagebreak \noi {\bf Theorem 5.}  The quantizations
given by the following formulas

\beq {\ooo} (q^n\,p^m)  =  \frac{1}{2^{\,n}} \sum_{i=0}^n \,
\qk^{\, n-i}\, \pk^{\, m}\,\qk^{\, i}, \label{o1}\eeq

\beq {\sss}(q^n\,p^m)  =  \lrb \ba{c} n+m \\
        n\ea\rrb ^{-1}\,
 \sum_{\sigma(\qk,\pk)\, \in\, \Omega_{n,m}^2}
\sigma(\qk,\pk) \label{s1} \eeq

\noi are the same. \vspace{1mm}

\noi {\sl Proof.} We will show that the right sides of the
formulas (\ref{o1}) and (\ref{s1}) have same coefficients in the
standard basis of $\hat {\cal U}({\cal H}_3).$ Firstly, let us
start with the following identities:
 \bea  & & \label{id1} \lrb\ba{c} n \\k \ea
\rrb + \lrb \ba{c} n \\ k-1\ea \rrb = \lrb \ba{c} n+1 \\k \ea \rrb,\quad k\leq n,
\vspace{10mm} \\
 & & \label{id2} \sum_{k=0}^n \lrb \ba{c} n \\k \ea \rrb\, \lrb \ba{c} k
\\ j \ea \rrb\, j\,! = 2^{n-j}\, n\,(n-1)\cdots (n-j+1),\quad j\in
\mathbb{N}_0\,. \eea

\noi The relation (\ref{e21}) is equivalent with  the following
relation \beq \pk\, \qk =\qk\, \pk + \hk,\label{r3} \eeq where, as
we introduced before, $\hk=-\, i\, \hbar\,\ik.$ \noi Then using
the relation (\ref{r3}), by induction one can show the following
identities.\vspace{3mm}

\noi {\bf Lemma 5.1.} $\forall\, m,\, n\in \bf N$  holds
\vspace{0mm} \bea \label{r4a} \mbox{(i1)}\ \ & &  \pk^m\, \qk =
\qk\, \pk^m +m\,
\pk^{m-1}\, \hk ,\vspace{4mm} \\
\label{r4b}\mbox{(i2)}\ \ & & \pk^m\, \qk^n = \sum_{k=0}^{n\wedge
m} \lrb \ba{c} n \\k \ea \rrb\, \lrb \ba{c} m
\\k \ea \rrb\, k\,!\ \qk^{\, n-k}\, \pk^{\, m-k}\, \hk^k\,. \eea
where $n\wedge m =\min (n,m)\,.$
 \vspace{2mm}

\noi Let us now find, using (\ref{id1}), (\ref{id2}) and  above
lemma, the coordinates of $ {\ooo}(q^{\, n}\, p^{\, m})$  in
standard basis of $\hat \htri \,. $ We have \bea {\ooo}(q^{\, n}\,
p^{\, m}) & = & \frac{1}{2^n} \sum_{k=0}^n \lrb \ba{c} n \\k \ea
\rrb\, \qk^{\, n-k}\, (\pk^{\, m}\, \qk^{\,
k}) \vspace{2mm} \nn\\
& = & \frac{1}{2^n} \sum_{k=0}^n \lrb \ba{c} n \\k \ea \rrb\,
\qk^{\, n-k} \lrb \sum_{j=0}^{k\wedge m} \lrb \ba{c} k \\j \ea
\rrb\, \lrb \ba{c} m \\j \ea \rrb\, j\,!\, \qk^{\, k-j}\, \pk^{\,
m-j}\, \hk^j\rrb \vspace{2mm}\nn \\
& = & \frac{1}{2^n} \sum_{j=0}^{n\wedge m} \lrb \sum_{k=0}^n \lrb
\ba{c} n \\k \ea \rrb\,   \lrb \ba{c} k \\j \ea \rrb\ j\,! \rrb
\lrb \ba{c} m \\j \ea \rrb\, j\,!\, \qk^{\, n-j}\, \pk^{\, m-j}\,
\hk^j \vspace{2mm}\nn \\
& = & \frac{1}{2^n} \sum_{j=0}^{n\wedge m} \lrb \ba{c} n
\\ j \ea \rrb\,   \lrb \ba{c} m \\j \ea \rrb\ 2^{\, n-j}\, j\,!\
\qk^{\, n-j}\, \pk^{\, m-j}\, \hk^j
\vspace{2mm}\nn \\
& = & \sum_{j=0}^{n\wedge m} \lrb \ba{c} n
\\j \ea \rrb\,   \lrb \ba{c} m \\j \ea \rrb\ 2^{\, -j}\, j\,!\
\qk^{\, n-j}\, \pk^{\, m-j}\, \hk^j \vspace{2mm}\label{r6} . \eea
\vspace{2mm}

\noi  If we introduce \beq \al_j^{n,m} =\lrb \ba{c} n \\j \ea
\rrb\, \lrb \ba{c} m \\j \ea \rrb\, \frac{j\, !\,}{2^j},
\label{r7} \eeq then we have \vspace{2mm}

\noi {\bf Lemma 5.2.} $\forall\, m,\, n\in \N, i\leq n,\ $ the
following relations hold \vspace{3mm}

\bi \i[(i1)]\quad $ \sum\limits_{k=0}^{n}\, \lrb \ba{c} n+m-k
\\m \ea \rrb = \lrb \ba{c} n+m+1
\\n
\ea \rrb\,,$ \vspace{4mm} \\
\i[(i2)] \quad $\sum\limits_{k=1}^{n}\, k\, \lrb \ba{c} n+m-k \\ m
\ea \rrb = \lrb \ba{c}
n+m+1\\n-1\ea\rrb\,,$\vspace{4mm}\\
\i[(i3)] \quad  $ \sum\limits_{k=i}^{n} \lrb \ba{c} k \\i \ea
\rrb\, \lrb \ba{c} m+k
\\m \ea \rrb = \lrb \ba{c}
n+m+1 \\ n \ea \rrb\, \lrb \ba{c} n \\ i \ea \rrb\,$
\begin{Large}$\frac{m\,+\,1}{m\,+\,1\,+\,i}$\end{Large}$\ ,$  \vspace{4mm}  \\
\i[(i4)] \quad $ \sum\limits_{k=0}^{n} \lrb \ba{c} n-k+m
\\m \ea \rrb\, \lrb\al_{j+1}^{n-k,m} + (n-k-j)\,
\al_{j}^{n-k,m}\rrb =$ \vspace{3mm}

\hspace{1cm}$\lrb \ba{c} m+n+1 \\m \ea \rrb \, \al_{j+1}^{n,m+1}.$
\ei

\vspace{2mm}

\noi {{\sl {Proof of Lemma 5.2.}}} (i1)-(i3) by induction. For
(i4), firstly we check the following relation,
\begin{small}\bea\label{al}\nn \al_{j+1}^{n-k,m} + \lrb n-k-j\rrb\,
\al_{j}^{n-k,m} = \frac{m+j+2}{2^{j+1}\,}\, \lrb \ba{c} n-k \\ j+1
\ea \rrb\, m \, (m-1)\,\dots \, (m-j+1),  \eea \end{small} \noi
then we have\vsd

\begin{small} \bea & & \sum_{k=0}^{n} \lrb \ba{c} n-k+m
\\m \ea \rrb\,  \lrb\al_{j+1}^{n-k,m} + (n-k-j)\, \al_{j}^{n-k,m}\rrb =
 \vspace{2mm} \nn\\
& & \hspace{10mm} \frac{m+j+2}{2^{j+1}\,}\,  m \, (m-1)\,\dots \,
(m-j+1)\, \sum_{k=0}^{n} \lrb \ba{c} n-k+m
\\m \ea \rrb\, \lrb \ba{c} n-k \\ j+1
\ea \rrb\, = \vspace{2mm}\nn \\
& &  \hspace{10mm} \frac{m+j+2}{2^{j+1}\, }\,  m \, (m-1)\,\dots
\, (m-j+1)\, \sum_{k=0}^{n} \lrb \ba{c} k+m
\\m \ea \rrb\, \lrb \ba{c} k \\ j+1
\ea \rrb\, = \
\vspace{2mm}\nn \\
&  & \hspace{10mm} \frac{m \, (m-1)\,\dots \, (m-j+1)\, (m+j+2)\,
(m+n+1)\, ! }{2^{j+1}\, \, (j+1)\, !\, m\, ! \, (n-j-1)\, ! }\, =
\nn \vsd\\
&  & \hspace{10mm} \lrb \ba{c} m+n+1 \\m \ea \rrb \,
\al_{j+1}^{n,m+1}\,. \nn \eea\end{small} \vspace{3mm}

\noi Let us show that ${\sss}(q^{\,n}\, p^{\,m})$ is equal to
(\ref{r6}). For $m=0$ and all $n$ it is clear that formula holds.
If we assume that it is true for some $m\in\mathbb{N}$ and
arbitrary $n\in {\mathbb{N}}_{\,0}$, then using Lemma 5.2., we
have \begin{small}
\bea \hspace{-.0cm}{\sss}(q^{\, n}& & \hspace{-.8cm}p^{\, m+1})\hspace{2mm} = \hspace{2mm}\lrb \ba{c} n+m+1\\
        n\ea\rrb ^{-1}\,
 \sum_{\sigma(\qk,\pk)\, \in\, \Omega_{n,m+1}^2}
\sigma(\qk,\pk)  \vspace{2mm} \nn\\
 & = &  \lrb \ba{c} n+m+1\\
        n\ea\rrb ^{-1}\, \sum_{n_i \geq  0
           \atop \sum_{i=0}^{m+1}n_i = m+1} \qk ^{n_0}\, (\pk\, \qk^{n_1})\,
           \cdots\, (\pk\, \qk^{n_{m+1}})
           \vspace{2mm}\nn \\
& = & \lrb \ba{c} n+m+1\\
        n\ea\rrb ^{-1}\, \sum_{n_0=0}^n \qk ^{n_0}\, \pk\, \sum_{n_i \geq  0
           \atop \sum_{i=1}^{m+1}n_i = m+1} \qk^{n_1}\, (\pk\, \qk^{n_{2}})
           \cdots\, (\pk\, \qk^{n_{m+1}})
\vspace{2mm}\nn \\
& = & \lrb \ba{c} n+m+1\\
        n\ea\rrb ^{-1}\, \sum_{n_0=0}^n \qk ^{n_0}\, \pk\, \lrb \ba{c} n-n_0+m\\
        m\ea\rrb\,  \vspace{2mm}\nn \\
&  & \times\,
 \sum_{j=0}^{(n-n_0)\wedge m}
        \al_j^{n-n_0,m}\,\,\qk^{n-n_0-j}\,\,\pk^{m-j}\,\, \hk^j
        \vspace{2mm}\nn \\
& = & \lrb \ba{c} n+m+1\\
        n\ea\rrb ^{-1}\, \sum_{n_0=0}^n \lrb \ba{c} n-n_0+m\\
        m\ea\rrb\,  \sum_{j=0}^{(n-n_0)\wedge m}
         (\, \al_j^{n-n_0,m}\,\, \qk^{n-j}\,\,\vspace{4mm}\nn \\
&  & \cdot\, \pk^{m+1-j}\,\, \hk^j  +
         (n-n_0-j)\, \al_j^{n-n_0,m}\,\,\qk^{n-j-1}\,\,\pk^{m-j}\,
         \hk^{j+1}\,)\vspace{7mm}\nn \\
& = &  \lrb \ba{c} n+m+1\\
        n\ea\rrb ^{-1}\, \sum_{j=0}^{n\wedge (m+1)}\ \, \sum_{n_0=0}^n\,
        \lrb \ba{c} n-n_0+m\\ m\ea\rrb\,
        \,  \vspace{7mm}\nn  \\
 &  & \times\,    (\, \al_{j+1}^{n-n_0,m} + (n-n_0-j)\, \al_j^{n-n_0,m}\,)\,
 \qk^{n-j-1}\,\pk^{m+1-j}\, \hk^{j}\vspace{4mm}\nn \\
 & = &    \sum_{j=0}^{n\wedge (m+1)}\, \al_{j}^{n,m+1}\,\,
 \qk^{n-j}\,\,
 \pk^{m+1-j}\, \hk^{j}\, .     \label{r7}\nn  \eea \end{small}\vsd

\noi This completes the proof of Theorem 5. \qed \vsp\vsp

\noi {\Large {\bf{ V \ \ Conclusion}}}\vsc

\noi The symmetrized product and symmetrized Poisson bracket we
have defined in a way that they in complete imitate the
appropriate operations in the c-number case. If there is some
equation for classical variables, then the same equation holds for
the quantum counterparts. Consequence of this is that there are,
we believe, neither algebraic nor Lie algebraic contradictions in
quantum mechanics based on these operations since no such
contradictions appear in classical mechanics. Moreover, we have
shown that classical and quantum mechanics are not just similar
from the point of view of algebra, Lie algebra and dynamical
equation, but have the same realizations of these important
features. \vsj

\noi Also, in the previous section we showed that our quantization
implies the same ordering rule as the Weyl quantization. This fact
and obvious symmetricity of our quantization explain why Weyl
quantization is so symmetric.\vsp\vsp

\noi {\Large {\bf{ Appendix A}}}\vsc

\noi With the help of: \bea\nn \vert q' \rangle \langle q' \vert =
\int \delta (q-q')\vert q \rangle \langle q \vert dq = \delta
(\hat q - q'), \eea and: \bea \nn \vert q'' \rangle \langle q'
\vert = e^{{1\over i\hbar} (q'' - q')\hat p} \vert q' \rangle
\langle q' \vert = e^{{1\over i\hbar} (q'' - q')\hat p } \cdot
\delta (\hat q - q'), \eea one immediately finds that the operator
corresponding to general pure state $\vert \psi \rangle = \int
\psi (q) \vert q \rangle dq$ can be expressed as:
$$
\vert \psi \rangle \langle \psi \vert = \int \int \psi (q) \psi ^*
(q') \vert q \rangle \langle q' \vert dq  dq' =
$$
\bea\nn =\int \int \psi (q) \psi ^* (q') e^{{1\over i\hbar} (q -
q')\hat p} \cdot \delta (\hat q - q') dq  dq'. \eea Since $\hat
\rho$ is $\sum _i w_i \vert \psi _i \rangle \langle \psi _i \vert
$, one concludes that all states of quantum mechanical system can
be expressed via operators of coordinate and momentum.\vsp\vsp

\end{document}